\documentclass{spie} 
\usepackage[]{graphicx}
\usepackage{url}
\usepackage{bm}
\usepackage{caption}
\usepackage{threeparttable}

\title{Simulation of the cosmic ray effects  for the LiteBIRD satellite  observing the CMB B-mode polarization  }

\author{Mayu Tominaga\supit{a,b}, Masahiro Tsujimoto\supit{a}, Samantha Lynn
Stever\supit{c,e},  Tommaso Ghigna\supit{d,e}, Hirokazu Ishino\supit{c}, Ken
Ebisawa\supit{a} for the LiteBIRD Joint Study Group
\skiplinehalf
\supit{a}Japan Aerospace Exploration Agency (JAXA), Institute of Space and Astronautical Science (ISAS), Sagamihara, Kanagawa, 252-5210 Japan;\\
\supit{b}Department of Astronomy, The University of Tokyo, Bunkyo-ku, Tokyo, Japan;\\
\supit{c}Department of Physics, Okayama University, 2-1-1, Okayama, Okayama, 700-8530 Japan; \\
\supit{d}Department of Physics, University of Oxford, Oxford, OX1, 2JD, United Kingdom;\\
\supit{e}Kavli Institute for the Physics and Mathematics of the Universe (Kavli IPMU),
The University of Tokyo, Kashiwa, 277-8583, Japan;\\
}
\authorinfo{E-mail: tominaga@ac.jaxa.jp}

\begin{document} 
\maketitle 

\begin{abstract}
 The LiteBIRD satellite is planned to be launched by JAXA in the late 2020s. Its main
 purpose is to observe the large-scale B-mode polarization in the Cosmic Microwave
 Background (CMB) anticipated from the Inflation theory. LiteBIRD will observe the sky
 for three years at the second Lagrangian point (L2) of the Sun-Earth system. Planck was
 the predecessor for observing the CMB at L2, and the onboard High Frequency Instrument
 (HFI) suffered contamination by glitches caused by the cosmic-ray (CR) hits. We
 consider the CR hits can also be a serious source of the systematic uncertainty for
 LiteBIRD. Thus, we have started a comprehensive end-to-end simulation study to assess
 impact of the CR hits for the LiteBIRD detectors. Here, we describe procedures to make
 maps and power spectra from the simulated time-ordered data, and present initial
 results. Our initial estimate is that $C_l^{BB}$ by CR is $\sim 2 \times
 10^{-6}~\mu$K$_{\mathrm{CMB}}^{2}$ in a one-year observation with 12 detectors assuming
 that the noise is 1~aW/$\sqrt{\mathrm{Hz}}$ for the differential mode of two detectors
 constituting a polarization pair.
\end{abstract}


\section{INTRODUCTION}\label{s1}
LiteBIRD is a satellite dedicated for observing the anisotropy of the linear
polarization of the Cosmic Microwave Background (CMB). It aims to detect the B-mode
signal at a large angular scale ($l<200$) with the sensitivity of the tensor-to-scalar
ratio $\Delta r \leq 0.001$ for constraining the theory of inflation. LiteBIRD has three
telescopes, (LFT, MFT, and HFT), covering 15 frequency bands over 34--448~GHz. At the
focal plane of each telescope, thousands of Transition Edge Sensor (TES) bolometers are
placed, cooled at 100~mK. LiteBIRD is planned to be launched in the late 2020s by JAXA
and is currently under design by an international collaboration among many institutions
in Europe, North America, and Japan. General descriptions can be found
elsewhere\cite{Sugai2020,Hazumi2019}.

LiteBIRD will scan the entire sky for three years from the second Lagrange point (L2) of
the Sun-Earth system.  It is becoming a popular destination among astronomical
satellites for providing a more thermally benign environment than near-Earth
orbits. However, L2 is also known to provide a more harsh environment in terms of
cosmic-ray (CR) radiation \cite{Ade2014}. CRs at L2 mainly consist of the Galactic
Cosmic Rays (GCR) and Solar Energetic Particles (SEP). GCR is a permanent component with
a hard spectrum, while SEP is an eruptive component with a soft spectrum associated with
solar flares. Two major effects are anticipated in response to CR radiation: total-dose
effects (TDE) and single event effects (SEE). Here, we consider SEE to the detector
caused by GCR. SEP events are eruptive and small in a fraction of time, thus the data
can be discarded during events similarly to Planck HFI\cite{Ade2014}.

Previous satellites suffered impacts on scientific products caused by SEE of the
detectors by GCR \cite{Catalano2014a}. It will be more serious for future satellites
with an increasingly high demand on instrument sensitivity. It is mandatory to assess
the impact of the CRs based on simulation at an early stage of the mission before fixing
the hardware design. This is currently underway for LiteBIRD. An early result is
described in a series of two articles. One is by Stever et al.\cite{Stever2021}, which
describes the overview of this study and covers the details of the simulation from the
CR spectrum to the Time-Ordered Data (TOD) and predicted $\Delta r$. This article covers
the details of the simulation from the TOD to sky maps and power spectrum.

We first start with a brief comparison with the Planck HFI (\S~\ref{s2}), which is the
predecessor to LiteBIRD, to predict the CR effects of LiteBIRD. We describe our
simulation in \S~\ref{s3} for the input data (\S~\ref{s3-1}) and the output products
(\S~\ref{s3-2}). In \S~\ref{s4}, we discuss its validity (\S~\ref{s4-1}) and scaleability
(\S~\ref{s4-2}), and discuss implications obtained through this exercise. We conclude in
\S~\ref{s5}.

\section{COMPARISON WITH PANCK HFI}\label{s2}
The High Frequency Instrument (HFI)\cite{Lamarre2010} on the Planck
satellite\cite{Ade2011} is the closest predecessor of LiteBIRD for observing the sky in
the 100--857~GHz band over a large angular scale at L2 using low-temperature detectors
cooled at 100~mK.  HFI had 20 spider-web bolometers (SWB) sensitive to thermal intensity
and 16 Polarization-Sensitive Bolometer pairs (PSB-a and PSB-b). The spider-like radio
absorbers were designed to minimize the cross section to CRs\cite{Holmes2008} and the
temperature rise by input energy was detected by NTD Germanium thermistors. Each
detector was situated on its own Si wafer die and was sensitive only to a single
frequency and a single linear polarization direction.

HFI achieved unprecedented sensitivity and demonstrated both the technical feasibility
and the superb scientific performance of low-temperature detectors operating at L2
\cite{Aghanim2020}. However, it also suffered contamination by CRs, which appeared as
numerous glitches of a short rise and slow decay profile in the TOD. Through the studies
of in-orbit data\cite{Ade2011,Ade2014} as well as ground testing\cite{Catalano2014a}, it
was found that the glitches are classified into three populations characterized by
different time scales and amplitudes, which were dubbed as short, long, and slow
glitches. The long glitches were the most numerous population, which were caused by CR
events hitting the Si die of the detector. The rapid propagation of thermal energy as
ballistic (athermal) phonons played a role in heat conduction. Short glitches were less
dominant, which were caused by CR events directly impacting the absorbers and
thermisters. Slow glitches were least frequent, which were considered to originate from
the part unique to PSB-a, as the population was only found in PSB-a. Eventually, these
glitches were removed by subtracting a template of each population in the time domain
during the ground pipeline processing.

\begin{figure}[hbtp]
\centering
 \includegraphics[keepaspectratio, width=0.9\columnwidth]{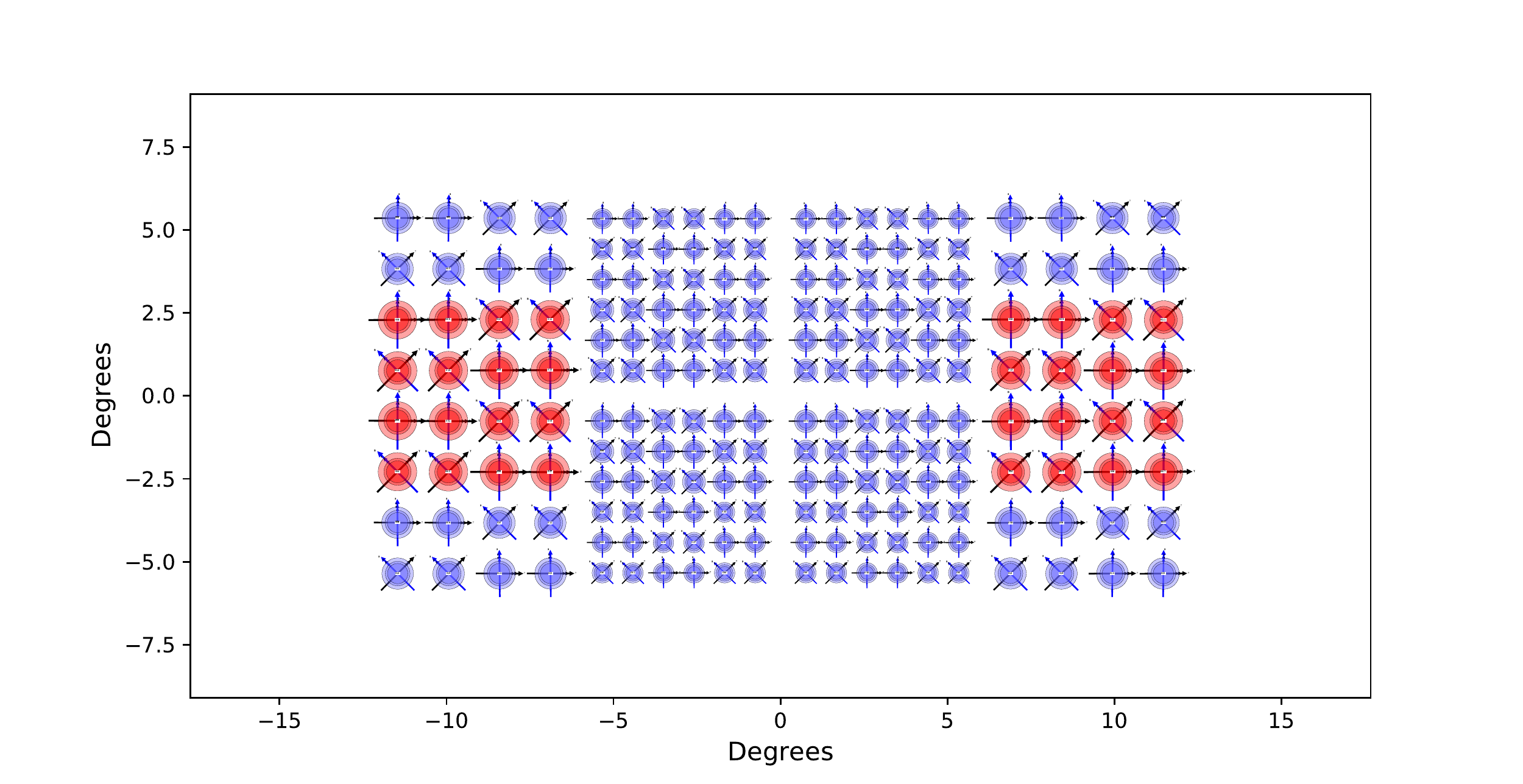}
 \caption{Focal plane layout of LFT. The circle represents the position of each pixel,
 which is sensitive to three bands (represented by circles proportional to beam size) in
 two linear polarization directions (by two arrows orthogonal to each other). The part
 used for the simulation in this study is the three pixels in the red area.}
 \label{f1-04}
\end{figure}

Accounting on some major design differences between LiteBIRD and Planck HFI, we can
predict the impact to some extent. The first difference is the number of detector
channels, in which the number increases by $\sim$10$^{2}$ times for LiteBIRD
(Table~\ref{t01}). A single detector has multiple channels sensitive to different
frequencies and directions of linear polarization. Many detectors are fabricated in
a single common Si wafer (Fig.~\ref{f1-04}). A single CR hit on a wafer will impact all
the detectors on the wafer, and cause common-mode noise among different frequency
bands and between two polarization angles. As the wafer area and volume are much larger
in LiteBIRD than Planck HFI, the event rate of the long glitches will be much higher but
the amplitude will be smaller for a larger heat capacity. These may make the noise
spectrum caused by long glitches behaves more like a Gaussian white noise.

\begin{threeparttable} 
 \centering
 \caption{Comparison of Planck HFI and LiteBIRD. For the wafer, LF-1 is used for
 LiteBIRD as a representative.}
 \label{t01}
 \begin{tabular}[hbtp]{lccccccccc}\hline\hline
  Satellite/  & \multicolumn{4}{c}{Thermometer}     & \multicolumn{3}{c}{Si wafer}\\
  Instrument  & Type   & $\tau$\tnote{a} & $f_{\mathrm{sample}}$\tnote{b} &
  Area\tnote{c}     & $C$\tnote{d}    & Det num\tnote{e} & Area\tnote{c} & Depth\tnote{f} & $C$\tnote{d} \\
              &        & (ms)   & (Hz) & (mm$^2$) & (pJ/K) &         & (cm$^{2}$) &
  (mm) & (nJ/K)\\
  \hline
  Plank / HFI & NTD Ge & 10     & 180 & 0.03     & 0.9    & 1       & 0.4-0.8 & 0.36 & 0.3 \\
  LiteBIRD    & TES    &  3     & 19  & 0.4      & 1.1    & 216     & 100 & 2.5 & 15\\
  \hline
 \end{tabular}
 \begin{tablenotes}[para,small]
  \item[a] Time constant of the detector, including thermo-electric feedback of a loop gain of 10 for TES.
  \item[b] Downlink data rate per detector channel.
  \item[c] Area of target. Surrounding parts included for LiteBIRD.
  \item[d] Heat capacitance of target.
  \item[e] Number of detector channels in a wafer excluding dark detectors.
  \item[f] Depth of target.
 \end{tablenotes}
\end{threeparttable}

The second difference is the data rate. Because of the increased number of detectors
albeit a similar telemetry bandpass to the ground stations, the data rate per detector
must be reduced for LiteBIRD to 19~Hz. In Planck HFI, the rate of 180~Hz for each
detector allowed to resolve CR glitches in time with a time constant of $>$10~ms and
remove them by template fitting in the time domain. In contrast, in LiteBIRD, the data
rate is slower by $\sim$10 times and the detector time constant is faster with the
thermo-electric feedback of TES by $\sim$3 times. It is thus impossible to resolve
glitches in time. Removal techniques similar to Planck HFI cannot be employed.

The third difference is the use of a Half-Wave Plate (HWP) in LiteBIRD. The target sky
signals through the telescope optics are modulated at a frequency of
4$f_{\mathrm{HWP}}$, where e.g., $f_{\mathrm{HWP}}=0.77$~Hz is the rotation speed of the
HWP for LFT. CR signals are not through the telescope optics, but are demodulated as if
they were in the data processing together with the target signals. Under such design
conditions, the strategy to suppress CR contamination is to spread the CR power flatly
over a wide range of frequencies and increase the signal-to-noise ratio at a particular
frequency of 4$f_{\mathrm{HWP}}$. It is not straightforward to assess all these effects
in analytical calculations, particularly when these are convolved with telescope
responses, satellite spins, and precessions. We, therefore, embarked on end-to-end
simulation from the CR spectrum at L2 to the power spectrum of three-year observations.

\section{RESULTS}\label{s3} 
\subsection{Input data}\label{s3-1} 
\begin{figure}[htbp]
 \begin{minipage}{0.5\hsize}
  \center
  \captionsetup{width=0.9\linewidth}
  \includegraphics[width=70mm]{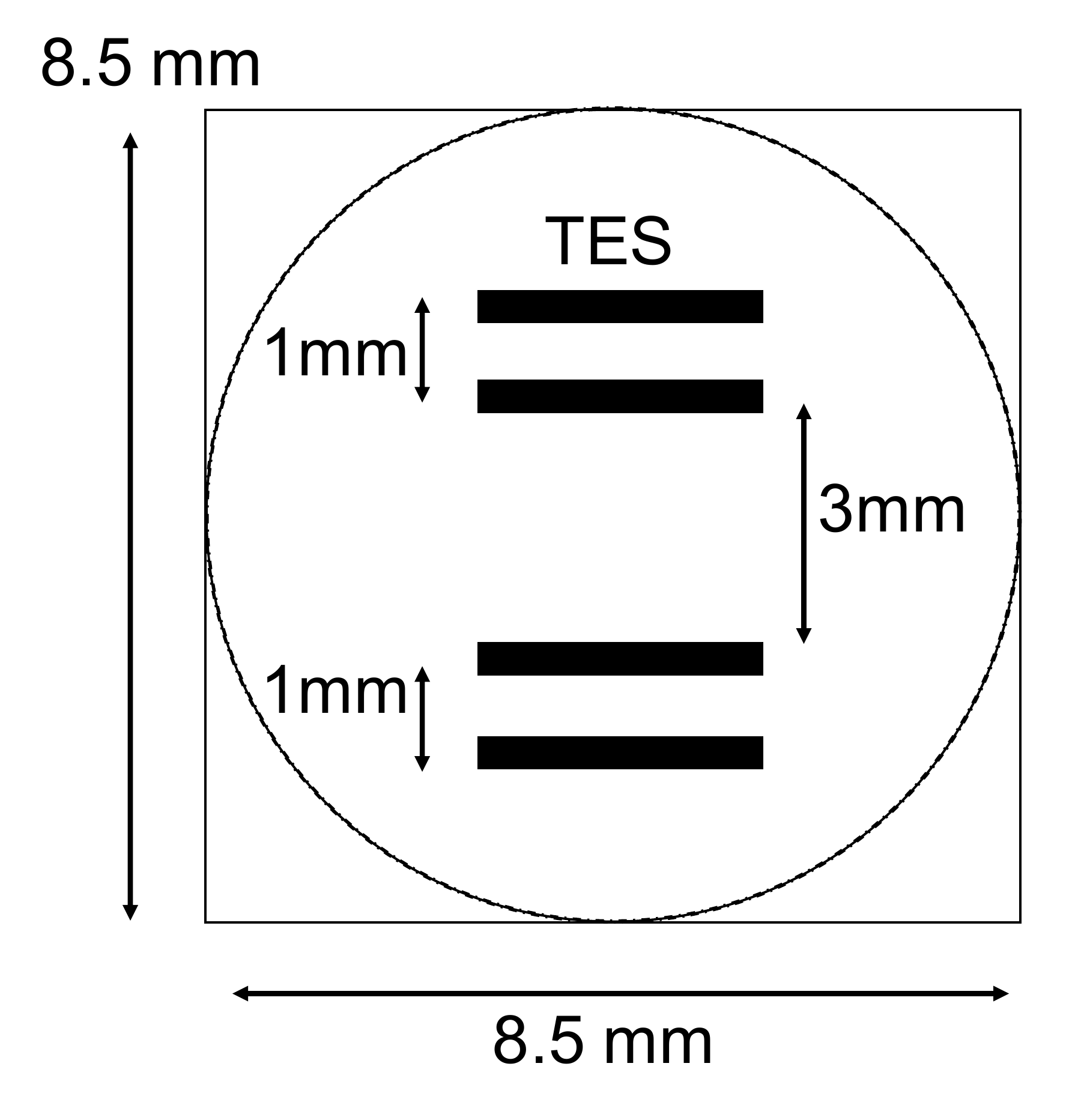}
  \caption{Position of TES in a pixel in the simulation. The circle represents the size
  of the lenslet for the pixel. The four parallel rectangles represent the TES. The pairs
  of top and bottom two are for the 78 and 100 GHz, respectively. Each pair reads two
  orthogonal directions of the linear polarization.}
  \label{f3-02}
 \end{minipage}
 \begin{minipage}{0.5\hsize}
  \center
  \captionsetup{width=0.9\linewidth}
  \includegraphics[width=70mm]{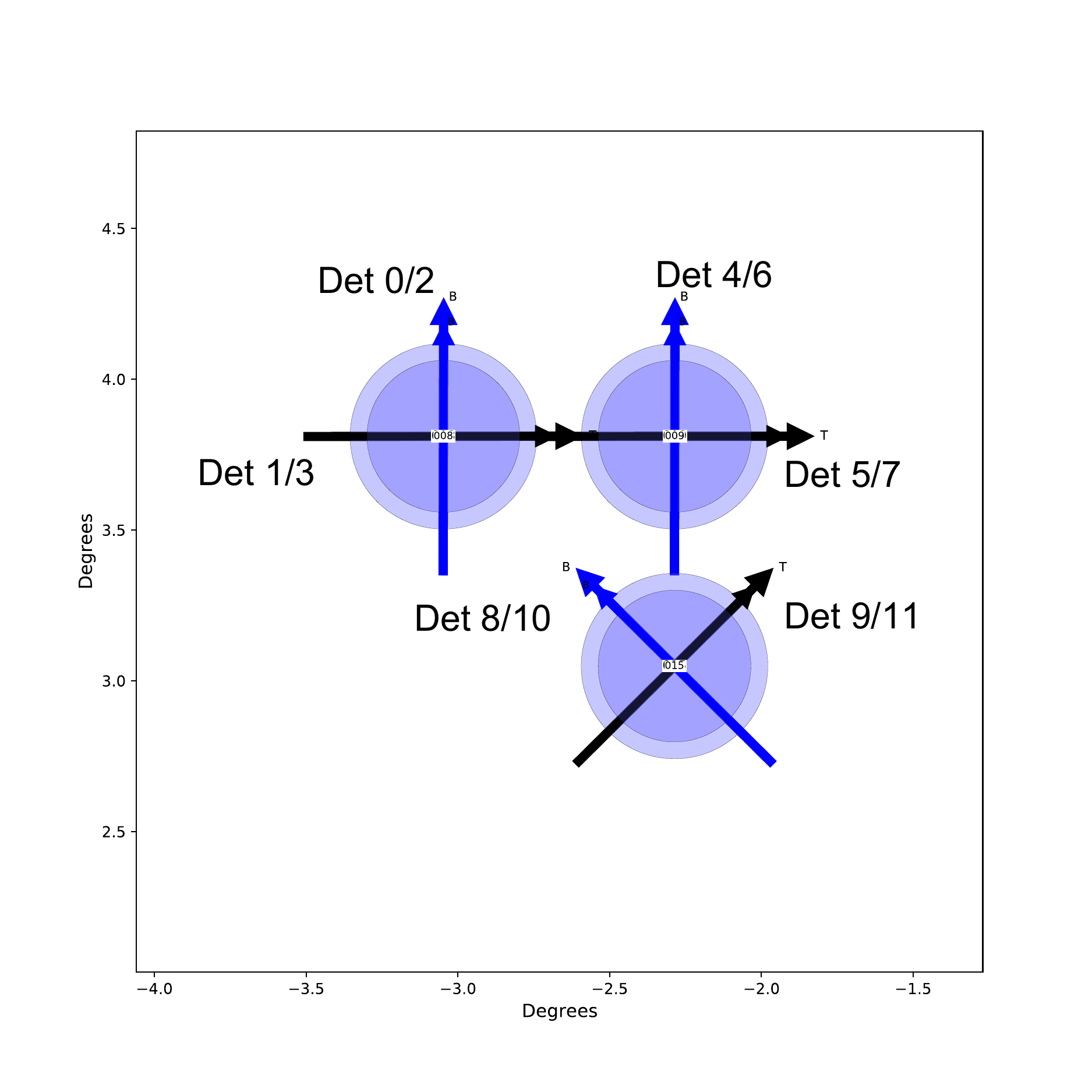}
  \caption{Part of the LF-1 wafer used in the simulation. Three pixels cover different
  positions of the sky separated by $\sim$0.75 arcdeg. Two polarization angles are shown
  with arrows, while two frequency bands are by circles proportional to the beam
  size. The assignment of the 12 channels is shown with the DetIDs.}
  \label{f3-01}
\end{minipage}
\end{figure}

We simulated TODs consisting only of the CR noise for a 90-minute length for 12 detector
channels on 3 pixels in a wafer (LF-1) of LFT (Fig.~\ref{f1-04}).  Each channel
represents one TES reading. In the simulation, one pixel stores four TES as shown in
Fig.~\ref{f3-02}.  Each TES reads antenna power of one of the two polarization
orientations in one of the two frequencies (78 and 100 GHz) in one of the three pixels
in the wafer as shown in Fig.~\ref{f3-01}.  The LF-1 wafer actually stores 6$\times$6
pixels of three frequencies of two polarization pairs with a total of 216 TES channels
without dark channels. For computational limitations, we simulated only 12 of them.

The LF-1 wafer has a size of 100$\times$100~mm$^{2}$ area and a depth of 2.5~mm made
with Si. The wafer is thermally anchored to the heat bath at the four sides. We
simulated the CR hits upon the LF-1 wafer following the GCR spectrum and flux obtained
with the PAMELA experiment\cite{Casolino2009, Picozza2007}. We expect $\sim$400~s$^{-1}$
hit per wafer. Each CR event deposits energy, which spreads over the entire wafer as
heat propagation. The average deposit energy is $\sim$1.3~MeV (0.2~pJ). The propagation
is calculated using a finite element method, and the temperature beneath the 12 TES
detector channels was derived as a function of time. Coupled differential equations
---one describing the thermal coupling between the TES and the Si wafer beneath the TES
and the other describing the electric circuit of TES under a constant bias current---
were solved. As a result, we obtained TOD made only with CR events in the unit of
power. The details of the simulation is given in Stever et al.\cite{Stever2021}.

\begin{figure}[hbtp]
 \centering
 \includegraphics[keepaspectratio, width=0.48\textwidth]{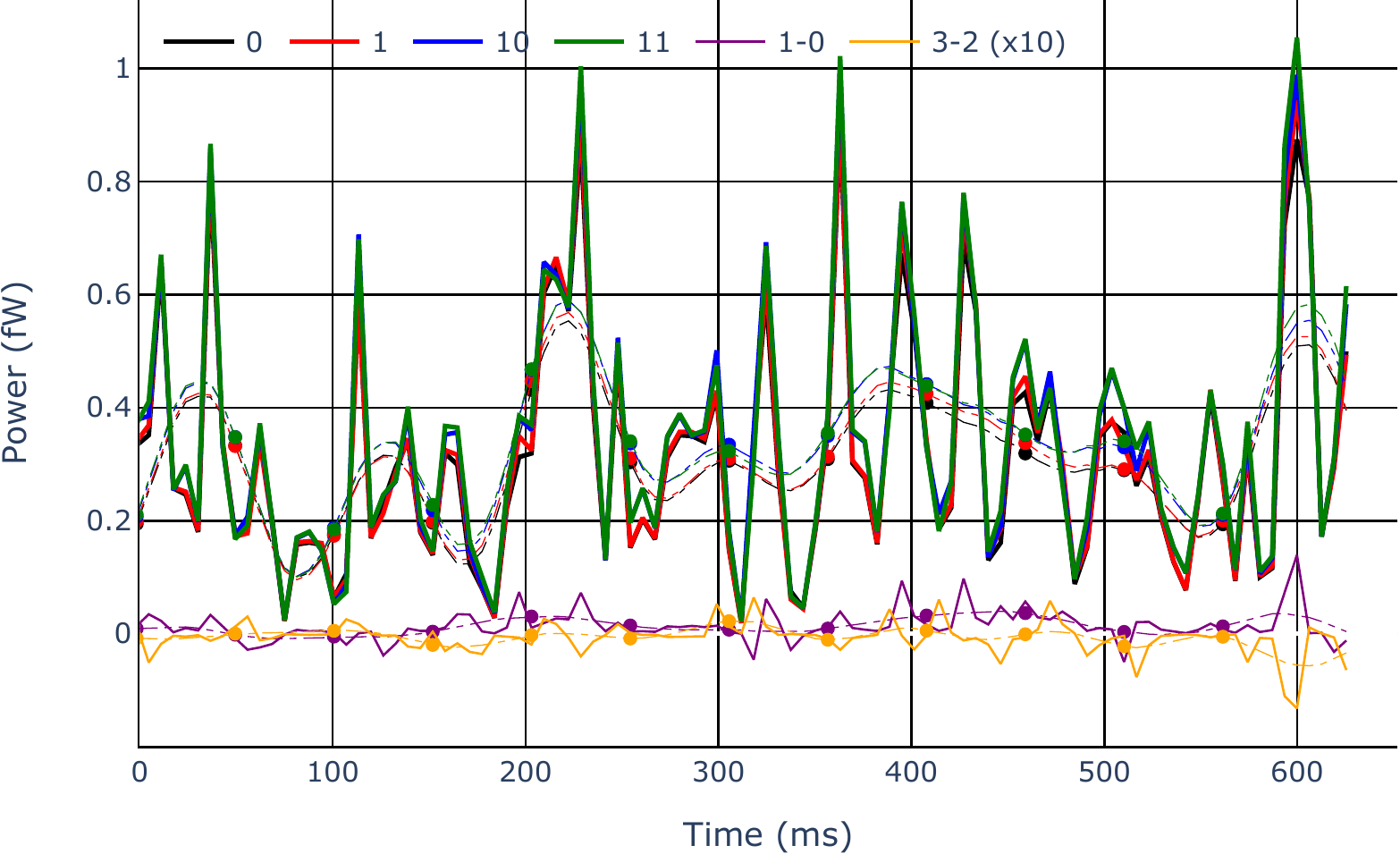}
 \includegraphics[keepaspectratio, width=0.48\textwidth]{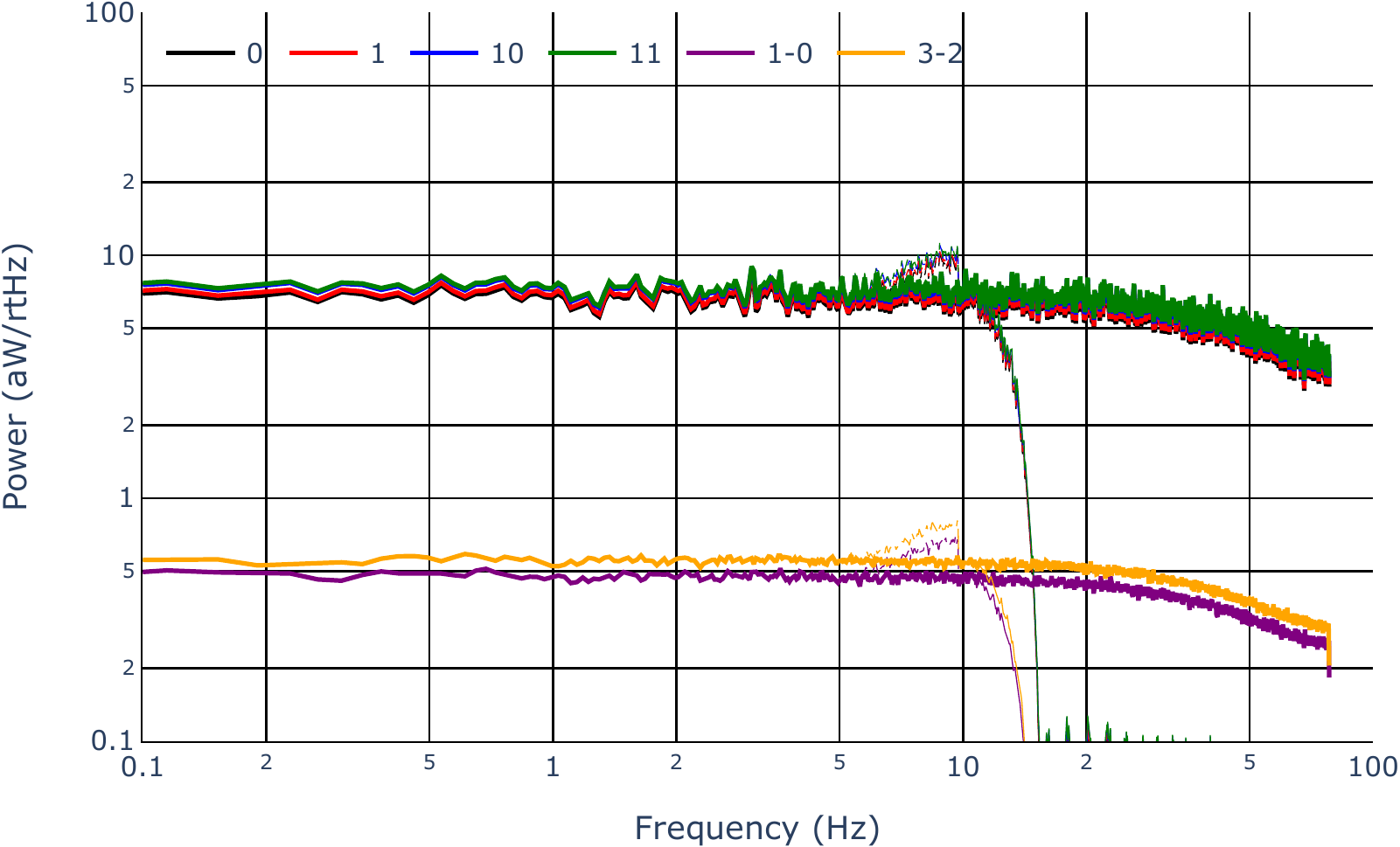}
 \caption{Input CR data in the time (left) and frequency (right) domains. The duration of
 time in the right is the same with two cycles of $4f_{\mathrm{HWP}}$ of LFT. Different
 channels are shown in different colors. Only four channels (det 0, 1, 10, 11) are shown
 for clarity. The difference between (0 and 1) and (2 and 3) is also shown. The 153~Hz
 data before decimation is shown by thick lines, while the 19~Hz date after decimation
 is by thin lines. The power can be converted to the equivalent thermo-dynamic
 temperature with a ratio of 0.2~aW $\sim$ 1 $\mu$K$_{\mathrm{CMB}}$.}
 \label{f4-04}
\end{figure}

The TOD in the time domain and its power spectrum are shown in Fig.~\ref{f4-04}. The TOD
is simulated at a time resolution of 153~Hz, which is decimated from the original 20~MHz
by a factor of 2$^{17}$ by multiple stages of the cascaded integrator and comb
filter. This is further decimated by a factor of 2$^{3}$ to 19.1~Hz to match with our
20~Hz telemetry bandpass using a finite impulse response digital filter. In the current
spacecraft design, the decimation by 2$^{17}$ is performed by the warm electronics using
FPGA, and the additional decimation by 2$^{3}$ is by the payload module digital processing unit
using CPU.

In the time domain, it is clearly seen that all 12 channels are strongly
correlated. This is what we expect for LiteBIRD. The LF-1 wafer has 216 TES channels,
all of which share a common Si wafer. At the temperature ($\sim$100~mK) of the Si wafer,
the heat generated by a CR propagates fast. The traveling speed is much faster than the
characteristic time scale of TES (a few ms) and the time resolution of the TOD (153~Hz)
regardless of whether the propagation is by diffusion or ballistic process. Therefore, a
single CR hit on the Si wafer affects all channels, which is almost synchronous in this
time resolution. In the frequency domain, all channels show a flat spectrum with a power
of $\sim$8.5~aW/$\sqrt{\mathrm{Hz}}$.

\subsection{Output maps and power spectra}\label{s3-2} 
We simulated maps using TODs of a one-year length to have a coverage of the entire sky
to calculate the angular power spectrum to the lowest scale of $l=2$. We produced
one-year length TODs in two different methods. One (method A) is to shuffle and
replicated the 90-minute TOD many times. This method is used when we evaluate the
effects related to the common-mode nature of the CR noise in time. We first divided the
TOD into ten pieces of a nine-minute length and selected one randomly 5840 times to make
it a one-year length. There are $10^{5840}$ combinations of such selection and a
sufficient number of realizations can be achieved in the simulation. All channels use
the simulated data at the same time in order to preserve the correlation among
them. This, however, produces artificial correlation in time within a given channel,
despite the fact that the TODs are shuffled. When we need to remove this effect, we made
the TODs with white noise randomly for the entire year for all the channels (method
B). The amplitude of the white noise was set to 1.0~aW/$\sqrt{\mathrm{Hz}}$. In making
$Q$ and $U$ maps, what matters is the differential mode of the TODs of two channels
constituting a polarization pair (1--0 in Fig.~\ref{f4-04} right), not the TOD of
individual two channels (0 or 1). When there is no correlation between the two channels,
the differential-mode TOD would have a white noise of an amplitude of $\sqrt{2} \times
8.5$~aW/$\sqrt{\mathrm{Hz}}$. When there is a strong correlation, as in the simulation,
the differential-mode TOD has an amplitude of $\sim$0.5~aW/$\sqrt{\mathrm{Hz}}$. The
simulation setup is not realistic enough to simulate the subtle difference between pair
channels, so we tentatively use 1.0~aW/$\sqrt{\mathrm{Hz}}$ as an amplitude and
interpret the result by scaling this value.

We used \texttt{TOAST}\footnote{See \url{https://toast-cmb.readthedocs.io/} for
details.}, which is a software framework for simulating and processing timestream data
collected by telescopes. The sky is represented by pixels with an equal area based on
the \texttt{HEALPix} \cite{Gorski2005} and its \texttt{python} implementation
\texttt{healpy} library \cite{Zonca2019}. We used the pixelization parameter
$N_{\mathrm{side}}=256$, which divides the entire sky with $N_{\mathrm{pix}} = 12
N_{\mathrm{side}}^{2}=786432$ pixels with an area corresponding to a square with a side
of 13.7~arcmin. The angular scale up to $l_{\mathrm{max}} = 767$ is obtained and the
number of independent components in the spherical harmonic function is
$N_{\mathrm{sph}}=(l_{\mathrm{max}}+2)(l_{\mathrm{max}}+1)/2=295296$.

\begin{figure}[hbtp]
\centering
 \includegraphics[keepaspectratio, width=\columnwidth]{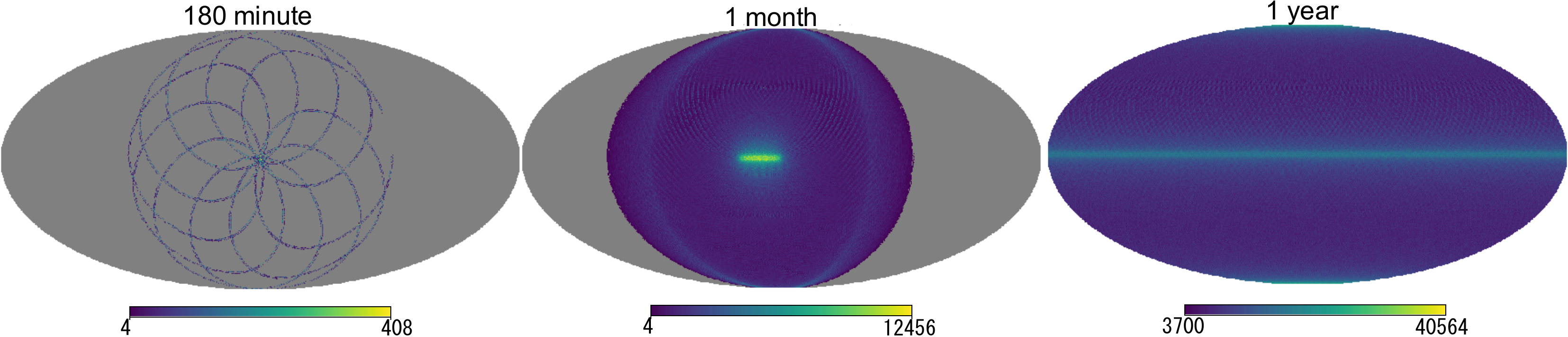}
 \caption{Hit maps over (left) 180 minutes, (middle) one month and (right) one year. The
 color bars indicate the number of visits for each pixel.}
 \label{f4-01}
\end{figure}

LiteBIRD scans the sky by a combination of a spin, a precession, and a rotation around
the Sun. The precession axis is along the anti-Sun direction, which moves at a speed of
360~deg year$^{-1}$. The precession rate is 192.348~min per rotation with an angle of 45
degrees. Along the precession axis, the satellite spins at a rate of 20 min per rotation
with an angle of 50 degrees. The resultant sky coverage is depicted in the hit maps
showing how many visits are made by detectors for each sky pixel in a given observation
time (Fig.~\ref{f4-01}).

At a given time $t$, the satellite is observing at a sky position $p(t)$ characterized
by the Stokes parameters $I(p)$, $Q(p)$, and $U(p)$. In LiteBIRD, the $U$ and $Q$
signals are further modulated by the HWP that is constantly rotating at a speed of
$\omega_{\mathrm{HWP}}$. The $i$'th detector signal at a time $t$ is then given by
\begin{equation}
 \label{e01}
 d_i(t) = I(p) + Q(p) \cos{2\psi_i} + U(p) \sin{2\psi_i} + n_i(t),
\end{equation}
where $n_i(t)$ is the noise and $\psi_i(t)$ is the modulation angle given by 
\begin{equation}
 \label{e02}
 \psi_i (t) =\psi_{0} (t) + \psi_{i}+2\omega_{\mathrm{HWP}} t,
\end{equation}
in which $\psi_{0} (t)$ is the relative angle of the focal plane to the sky, $\psi_{i}$
is the relative angle of the $i$'th detector to the focal plane. The modulation by HWP
is $4 f_{\mathrm{HWP}} = 4\omega_{\mathrm{HWP}}/2\pi \sim$3.1~Hz for LFT. In the matrix form,
\begin{equation}
 \label{e03}
 \bm{d} = \bm{P}(t,p) \bm{m}(p) + \bm{n},
\end{equation}
where $\bm{m}(p)$ is a vector of the three Stokes parameters. The map making is a process to
derive $\bm{m}(p)$ from observed $\bm{d}$. The simplest approach is to minimize 
\begin{equation}
 \label{e04}
 \chi^2=(\bm{d}-\bm{Pm})\bm{C}^{-1}_{\rm{n}}(\bm{d}-\bm{Pm}),
\end{equation}
where $\bm{C}_{\rm{n}}=\langle\bm{n}\bm{n}^T\rangle$ is the noise covariance, and obtain
the most plausible value of $\bm{m}(p)$ as
\begin{equation}
 \label{e05}
 \tilde{\bm{m}}=(\bm{P}^T \bm{C}_{\rm{n}}^{-1} P)^{-1} P^T \bm{C}_{\rm{n}}^{-1} \bm{d}.
\end{equation}

In making maps with \texttt{TOAST} from the simulated TOD (Fig.~\ref{f4-04}), we
developed a new module so that we can feed noise as a time series. The map making was
made separately for each day using multiprocessing and co-added in the map domain to
make a one-year map. In this manner, we can complete the map making in a few hours using
a personal computer equipped with 12 CPUs with a 3~GHz clock. We do not use the
destriping technique developed mainly to mitigate long-term fluctuations such as 1/$f$
detector noise. More sophisticated algorithm can be developed in the future by making
the full use of the features of the CR noise.

\begin{figure}[hbtp]
\begin{tabular}{ccc}
      \begin{minipage}[t]{0.33\hsize}
    \centering
    \includegraphics[keepaspectratio, width=\columnwidth]{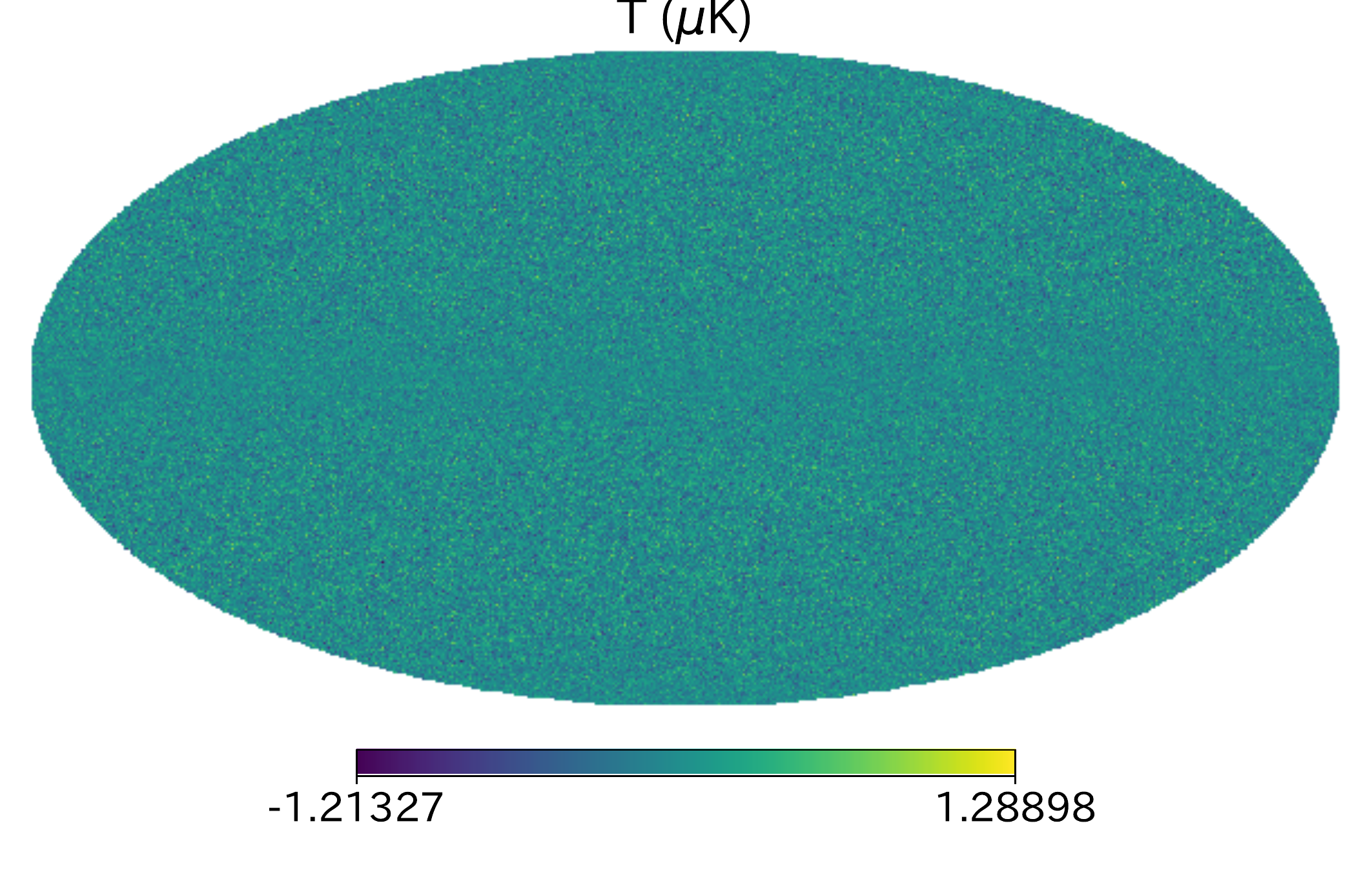}
    \end{minipage}
      \begin{minipage}[t]{0.33\hsize}
    \centering
    \includegraphics[keepaspectratio, width=\columnwidth]{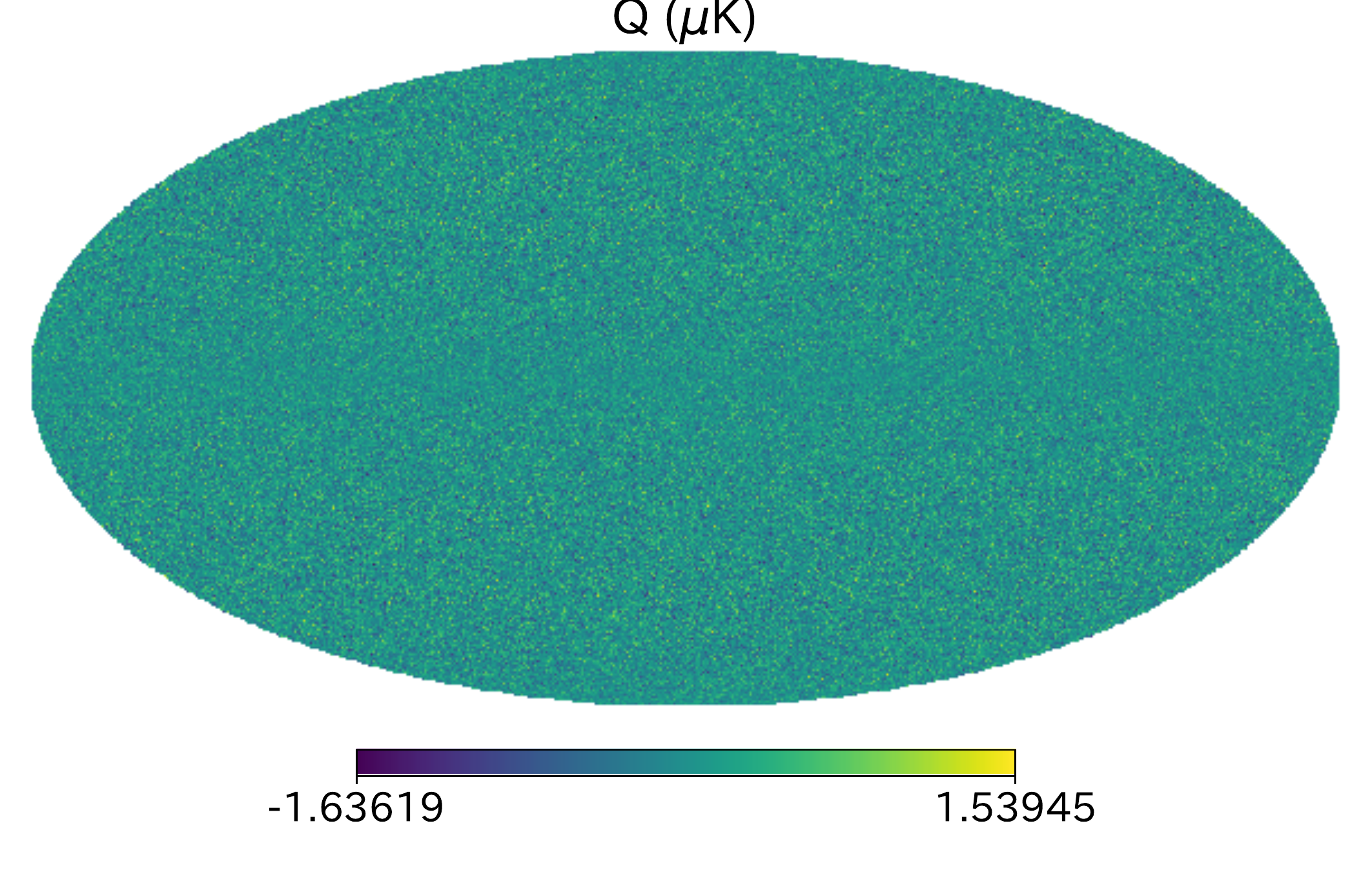}
    \end{minipage}
      \begin{minipage}[t]{0.33\hsize}
    \centering
    \includegraphics[keepaspectratio, width=\columnwidth]{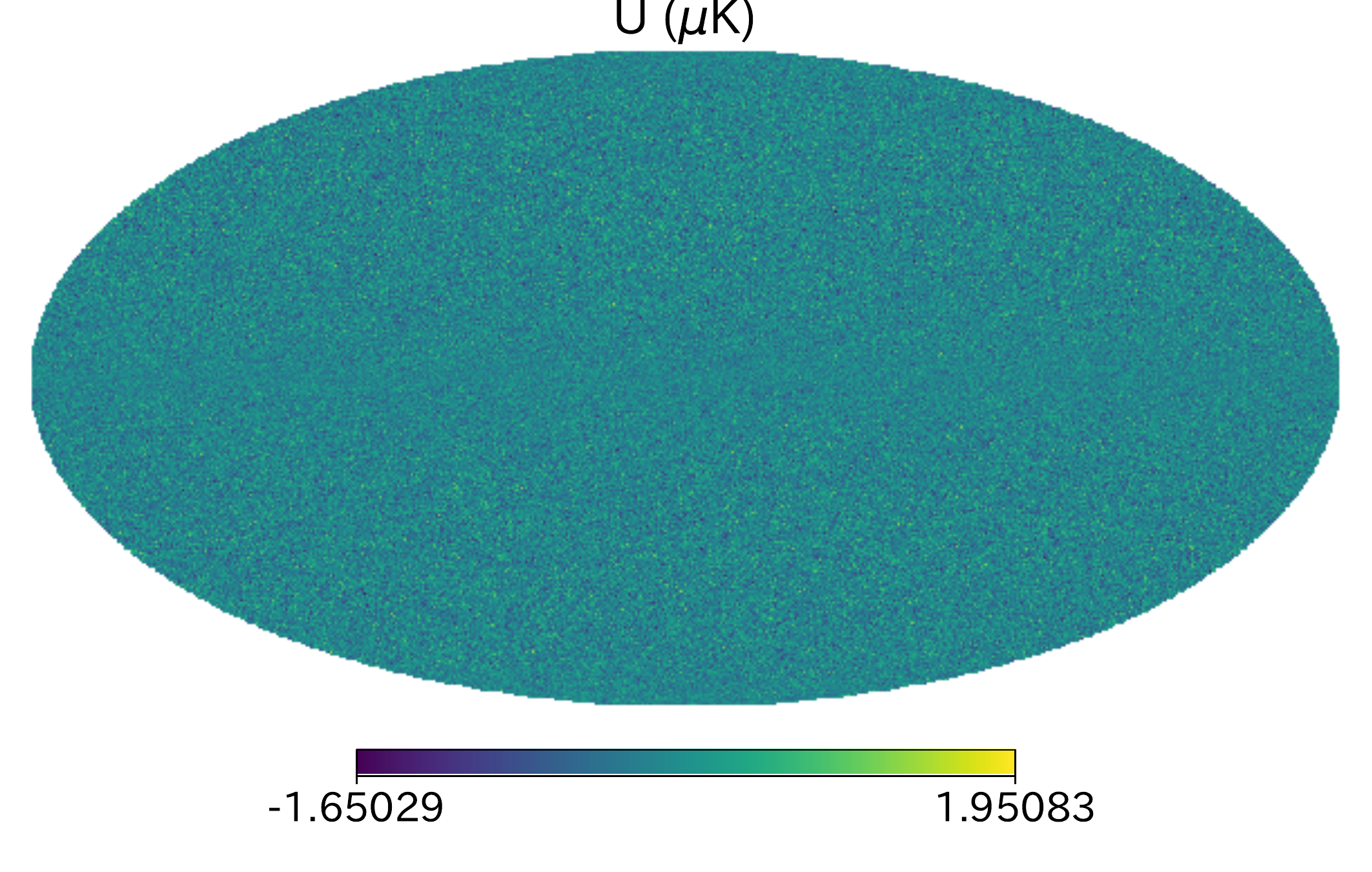}
    \end{minipage}
    \end{tabular}
     \caption{Simulated maps in the equatorial coordinate in the Mollweide projection:
 $T$ (left), $Q$ (middle), and $U$ (right) in the unit of $\mu$K$_{\mathrm{CMB}}$.}
 \label{f4-02}
\end{figure}

Fig.\ref{f4-02} shows the resultant $T=I$, $Q$ and $U$ maps of one year made with the
method B. The statistics of the maps are shown in Table~\ref{t4-01}. The monopole
component is removed from the $T$ map, but not from the $Q$ and $U$ maps. Still, the
mean value is small enough in the $Q$ and $U$ maps, indicating that they are made mostly
from the differences of polarization pair channels that are balanced with each other.
Finally, we calculated the power spectra using the \texttt{anafast} program of the
\texttt{healpy} library. The program separates the $E$ and $B$ components and calculates
the auto power spectra of $C_l^{TT}$, $C_l^{EE}$, and $C_l^{BB}$ as well as the cross
power spectra of $C_l^{TE}$, $C_l^{TB}$, and $C_l^{EB}$. We show $C_l^{EE}$ and
$C_l^{BB}$ relevant for the B-mode polarization detection.

\begin{table}[hbtp]
\centering
 \caption{Statistics of the maps.}
 \label{t4-01}
 \begin{tabular}[hbtp]{cccccc}\hline\hline
  Map & Mean ($\mu$K$_{\mathrm{CMB}}$) & Min ($\mu$K$_{\mathrm{CMB}}$) & Max
	      ($\mu$K$_{\mathrm{CMB}}$) & RMS ($\mu$K$_{\mathrm{CMB}}$) & $C_{l,est}$($\mu$K$_{\mathrm{CMB}}^2$) \\\hline 
  $T$ & --1.4$\times 10^{-4}$ & --1.2  & 1.2 & 0.23  & 2.4$\times 10^{-6}$\\
  $Q$ &	--2.6$\times 10^{-4}$  & --1.8 & 1.7 & 0.33 & 4.8$\times 10^{-6}$\\
  $U$ &	--1.3$\times 10^{-4}$  & --1.9 & 1.6 & 0.33 & 4.8$\times 10^{-6}$\\
  \hline
 \end{tabular}
\end{table}

\section{DISCUSSION}\label{s4}
\subsection{Confirmation of results}\label{s4-1}
\subsubsection{Realizations}
\begin{figure}[hbtp]
\begin{tabular}{cc}
      \begin{minipage}[t]{0.45\hsize}
    \centering
    \includegraphics[keepaspectratio, width=\columnwidth]{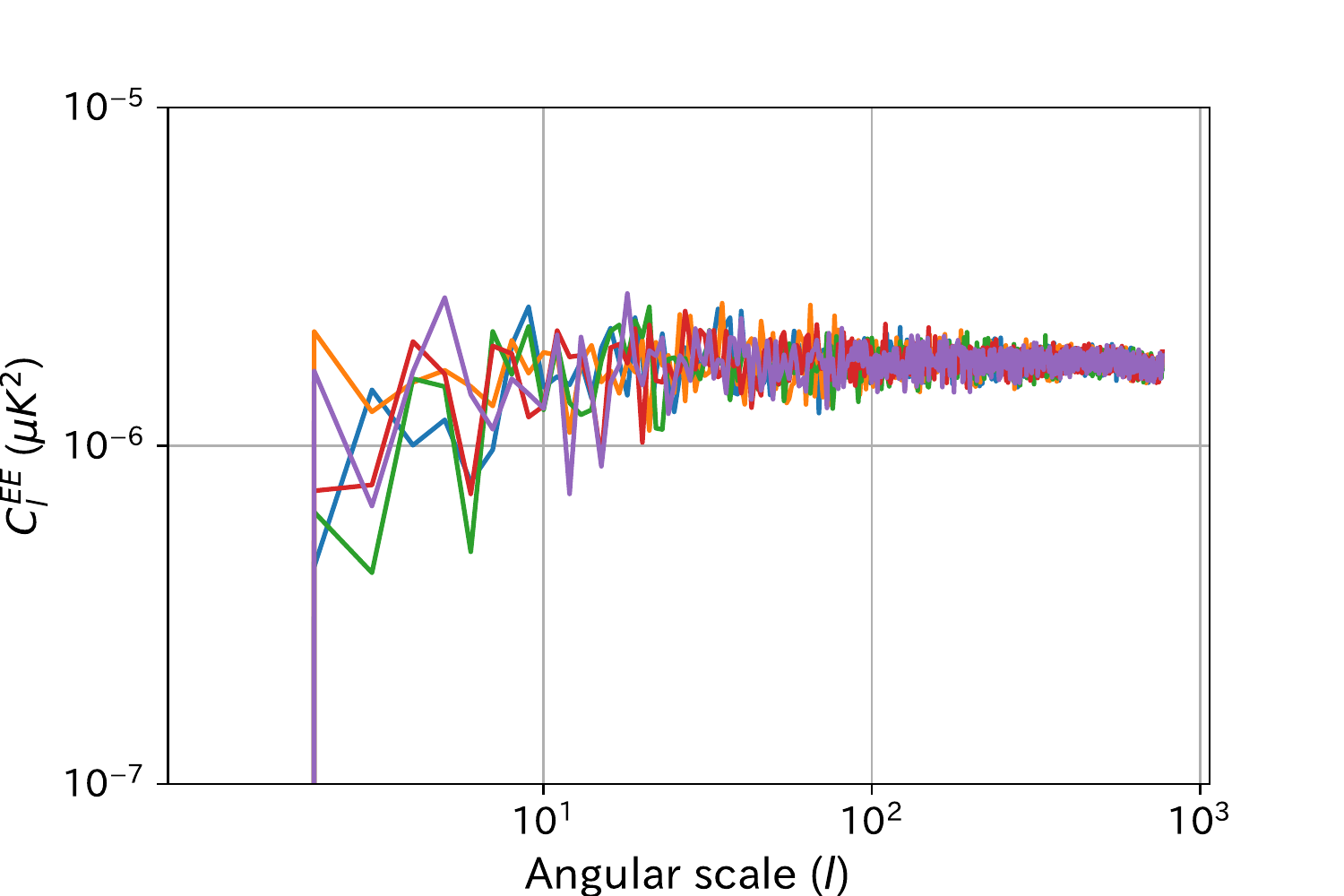}
    \end{minipage}&
      \begin{minipage}[t]{0.45\hsize}
    \centering
    \includegraphics[keepaspectratio, width=\columnwidth]{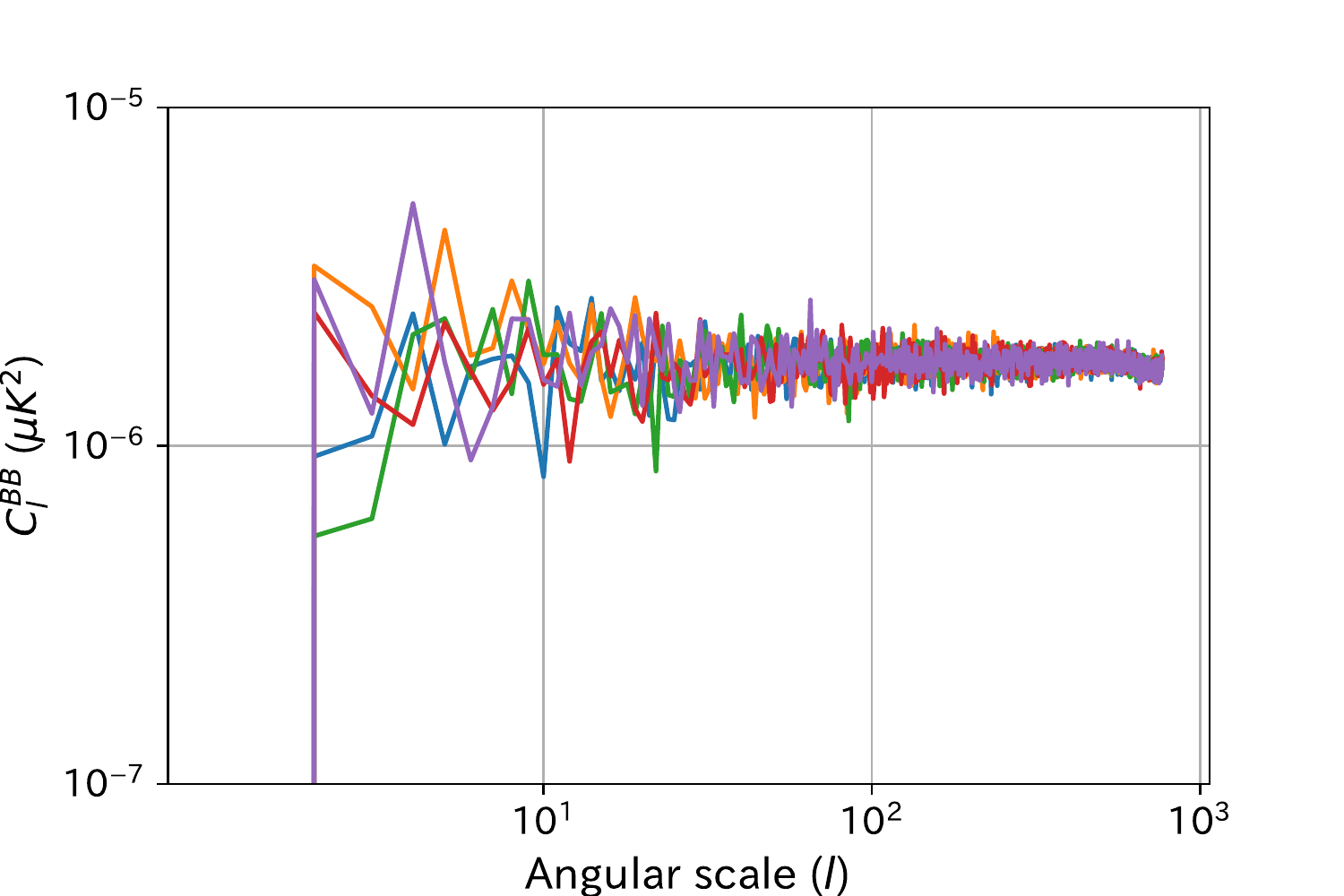}
    \end{minipage}
\end{tabular}
\caption{Power spectra of (left) $C_l^{EE}$ and (right) $C_l^{BB}$ for five different
 realizations of the one-year map simulation.}
\label{f4-12}
\end{figure}

We first check how much fluctuation is expected in the simulation. We repeated the
map-making with five different realizations of one-year TODs made the method B and
compared the results in Fig~\ref{f4-12}. The power spectra values in both $C_l^{EE}$ and
$C_l^{BB}$ fluctuates by a factor of a 2--3 at low-$l$, which provides an estimate of
the uncertainty in the simulation.

\subsubsection{Values of output products}
We next check if the values of the output products are consistent with the input TOD by
an order estimate. First, the TOD has a power of
$\sim$5.0~$\mu$K$_{\mathrm{CMB}}$/$\sqrt{\mathrm{Hz}}$ at $4f_{\mathrm{HWP}}$ for the
differential mode (1--0 or 3--2). This is relevant for estimating the noise in the $Q$
and $U$ maps because the two channels read two orthogonal directions at the same
position at the same time in the same band (Fig.~\ref{f3-01}), hence their common mode
is attributed more to the $I$ component and the differential mode to the $Q$ or $U$
component in Eqn.~\ref{e01}.
In a one year observation, each sky pixel is exposed for
\begin{math}
 T_{\rm{pix}}=1\times365\times24\times60\times60/N_{\rm{pix}} \sim 40~\mathrm{s}
\end{math}
on average. Therefore, the RMS value of the map is expected to be 
\begin{math}
 \sim 5.0/\sqrt{2T_{\rm{pix}}N_{\mathrm{det}}} \sim 0.4~\rm{\mu K_{\mathrm{CMB}}}
\end{math}
(Table.~\ref{t4-01}).
Here, $N_{\mathrm{det}}=3$ is the number of detector pixels (see discussion
below in \S~\ref{s4-2-2}). This is consistent with the values in Table~\ref{t4-01} for
the $Q$ and $U$ maps.
From the RMS of these maps, we can estimate the value of the coefficients of the spherical harmonic
oscillator function as
\begin{math}
 |a_{lm,\mathrm{est}}| \sim \mathrm{RMS} \sqrt{4\pi/N_{\mathrm{sph}}}.
\end{math}
From this, we can further estimate the value of angular power spectrum as 
\begin{math}
 C_{l,\mathrm{est}} \sim |a_{lm,\mathrm{est}}|^2.
\end{math}
The esitimated $C_{l,\mathrm{est}}$ is given in Table~\ref{t4-01}. The values for $Q$
and $U$ maps can be compared to $C_l^{EE}$ and $C_l^{BB}$ (Fig.~\ref{f4-12}) assuming
that the fluctuation of $EE$ and $BB$ are equivalent to the that of $Q$ and $U$. They
agree within a factor of a few.

\subsubsection{Distribution of output products}
\begin{figure}[hbtp]
 \centering
 \includegraphics[keepaspectratio, width=1.0\columnwidth]{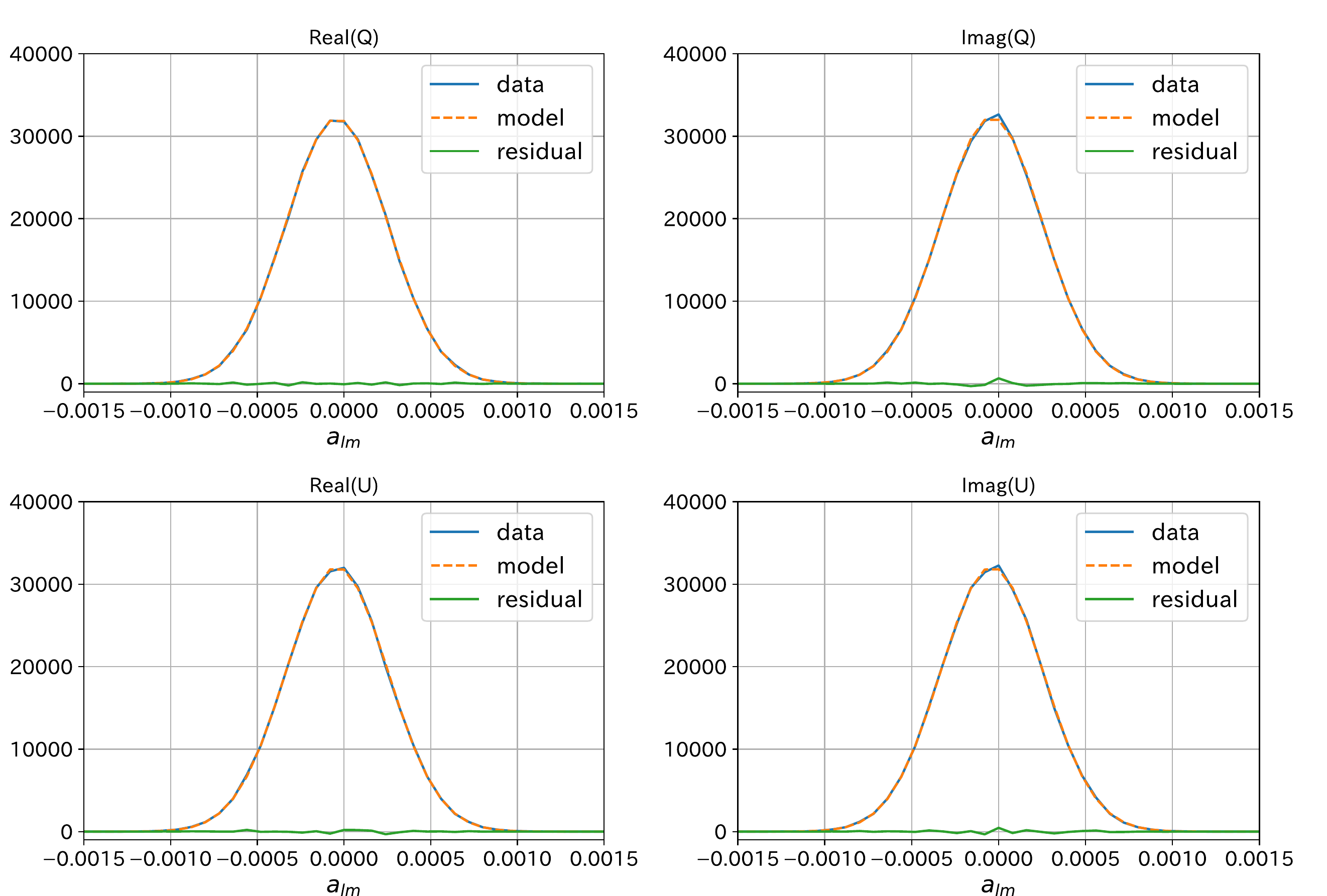}
 \caption{Distribution of $a_{lm}$ values of $Q$ (top) and $U$ (bottom) maps separately
 for the real (left) and imaginary (right) parts. The simulation results made with the
 method A are shown in blue, which is fitted with a single Gaussian model in orange. The
 residuals to the fit are in green.}
 \label{f4-05}
\end{figure}

We finally check that the $Q$ and $U$ maps by CR can be approximated as Gaussian, which
was implicitly assumed when adopting the method B. We made maps using the TODs made with
the method A, expecting that the distribution of the $a_{lm}$ follows the Gaussian. The
distribution is shown in Fig.~\ref{f4-05} separately for the real and imaginary parts
and for $Q$ and $U$ maps, which is fitted very well with a single Gaussian function.

\subsection{Dependencies}\label{s4-2}
\subsubsection{On observation time}\label{s4-2-1}
\begin{figure}[hbtp]
 \centering
 \includegraphics[keepaspectratio,width=\columnwidth]{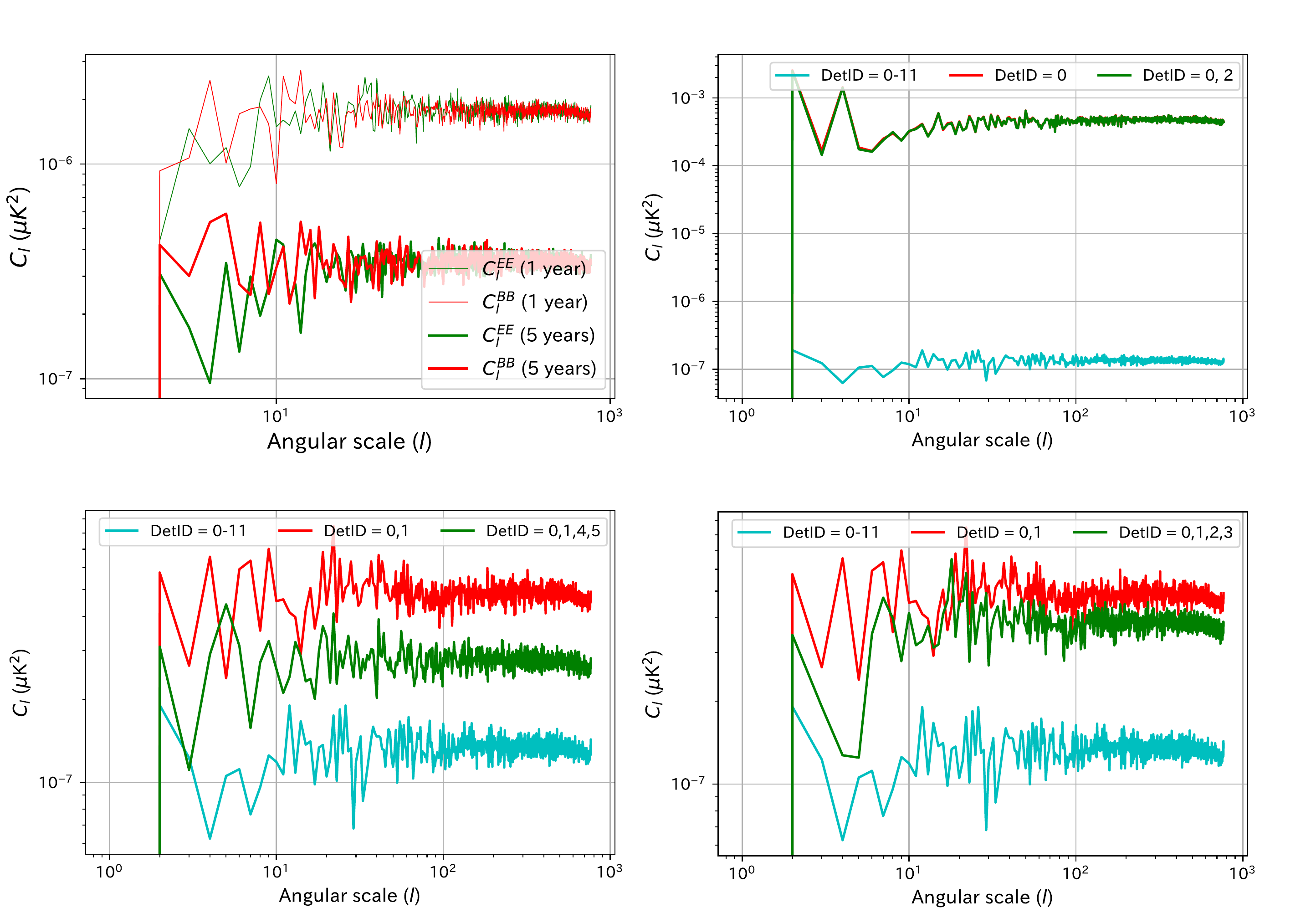}
 \caption{(Top left) Comparison of power spectra $C_l^{EE}$ (green) and $C_l^{BB}$ (red)
 for one-year (thin lines) and five-year (thick lines) observations using the method
 B. (Others) Comparison of power spectrum $C_l^{BB}$ for different subsets of detector
 channels using the method A.}
 \label{f4-08}
\end{figure}

We first investigate the dependence of the angular power spectra on the observation
time. As the TOD made with the method B is not correlated in the direction of time, we
expect that the $C_{l}$ value will decrease proportionally to the inverse of the total
integration time. We combined all the five one-year maps used in \S~\ref{s4-1} to
simulate a five-year observation and compared the result with the one obtained with one
year. The result is shown in Fig.~\ref{f4-08}. The $C_{l}$ value in the five-year
observation was reduced from that in the one-year observation by a factor of $\sim$5
both in EE and BB as expected.

\subsubsection{On the number of detector channels}\label{s4-2-2}
We next investigate the dependence on the number of detector channels. In this case, the
situation is more complex, as all the detectors are strongly correlated
(Fig.~\ref{f4-04}) in time. Some observe different polarization angle, sky position, or
frequency band. To investigate the dependence, we used the TODs made with the method A
and generated maps and power spectra from a subset of the 12 detector channels
(Table~\ref{t4-03}). The detector identification (DetID) follows Fig.~\ref{f3-01}. The
mean and the standard deviation of the TOD is slightly different among the 12 channels
reflecting the fact that they are located in a different location in the LF-1 wafer;
i.e., in the different distances from the thermal anchors on the four sides of the
wafer. We normalized the TOD to remove the effects by this difference so that we can
focus only on the dependence on the number of detectors.

\begin{table}[hbtp]
\centering
 \caption{Mean and standard deviation over $l$ of $C_l^{BB}$ for subsets of detectors
 made with the method A.}
 \label{t4-03}
 \begin{tabular}[t]{lccccc}\hline\hline
  Subset & Polarization & Pixel & Band & Mean ($\mu$K$_{\mathrm{CMB}}^{2}$) & Standard
		      deviation ($\mu$K$_{\mathrm{CMB}}^{2}$)\\
  \hline 
  0 & 1 &1  & 1 & 5.1$\times10^{-4}$ & 9.5$\times10^{-5}$\\
  0,2 & 1 & 1 & 2 & 4.7$\times10^{-4}$ & 9.3$\times10^{-5}$\\
  0,1 & 2 & 1 & 1 & 4.8$\times10^{-7}$ & 4.6$\times10^{-8}$\\
  0,1,2,3 & 2 & 1 & 2 & 3.8$\times10^{-7}$ & 3.7$\times10^{-8}$\\
  0,1,4,5 & 2 & 2 & 2 & 2.7$\times10^{-7}$ &2.4$\times10^{-8}$\\
  0--11 & 2 & 3 & 2 &1.3$\times10^{-7}$ & 1.3$\times10^{-8}$\\\hline
 \end{tabular}
\end{table}

First, we examine the angular power spectrum made only with DetID$=$0
(Fig.~\ref{f4-08}). The mean value of $C_l^{BB}$ is significantly elevated
(Table~\ref{t4-03}). This is because a large part of the CR noise is attributed to $Q$
and $U$ in Eqn.~\ref{e01}. In contrast, when we use both orientations (DetID$=$0, 1),
the value decreases as the dominant common-mode CR power is attributed preferentially to
$I$.

Second, we compare the result with DetID$=$(0, 1) and DetID$=$(0, 1, 4, 5) in
Fig.~\ref{f4-08} and Table~\ref{t4-03}. In the latter set, the DetID$=$(0, 1) and (4, 5)
pairs observe different positions of the sky at the same time. For a given position of
the sky, the two pairs observe at different times, hence are not correlated. Therefore,
the mean $C_l^{BB}$ value is reduced proportionally to the inverse of the pair number.

Third, we compare the result with DetID$=$(0, 1) and DetID$=$(0, 1, 2, 3) in
Fig.~\ref{f4-08} and Table~\ref{t4-03}. In the latter set, the DetID$=$(0, 1) and (2, 3)
pairs observe the same position of the sky at the same time. The two pair differences
0--1 and 2--3 are not correlated as much as individual channels (Fig.~\ref{f4-04}a),
thus some additional information can be obtained by adding another pair. Therefore, the
mean $C_l^{BB}$ value is reduced to some extent.

Finally, we compare the results using all detector channels DetID $=$ 0--11 with the
others. The mean $C_l^{BB}$ value is reduced by $\sim$3 times from DetID$=$(0, 1, 2, 3),
suggesting that the power spectrum increases proportionally to the inverse of the number
of sky positions observed at the same time.

\subsection{Implications}\label{s4-3}
In \S~\ref{s4-1}, we confirmed that (i) the CR noise is nearly Gaussian in the $Q$ and
$U$ map domain, and (ii) the values in the maps and the power spectra generated by
\texttt{TOAST} are consistent with the input TOD. In \S~\ref{s4-2}, we found that the
power spectrum $C_l^{BB}$ by CR noise is reduced inverse-proportionally when we increase
the integration time and the number of sky positions observed at the same time. If we
assume that the input CR power is 1~aW/$\sqrt{\mathrm{Hz}}$ for the differential mode of
two channels constituting a polarization pair, the expected $C_l^{\mathrm{BB}} \sim
2\times 10^{-6}~\mu$K$_{\mathrm{CMB}}^{2}$ for one year observation with three pixels
(Fig.~\ref{f4-08}a). This value scales with the exposure time and the number of pixels.
If this is the case, the impact of the CR upon the B-mode measurement of an order of
$\approx$10$^{-6} \mu$K$_{\mathrm{CMB}}^{2}$ would be at a manageable level, if not
negligible.

\begin{enumerate}
 \setlength{\itemsep}{-1mm}
 \item The assumption of 1~aW/$\sqrt{\mathrm{Hz}}$ for the differential mode power of
       the paired polarization channels needs to be verified. This depends on the detailed
       design of the detectors, which is not included in the simulation yet.
 \item The TOD of all detector channels are strongly coupled. This will bring a
       common-mode noise common among multiple frequency bands observed with the same
       detector pixel and pose a new challenge in separating CMB component from others
       in the spectrum.
 \item Two sky positions at a fixed separation should be strongly correlated, as they
       are observed at the same time with a given pair of pixels. This is not evident in
       our result (Fig.~\ref{f4-08}) with only three pixels, but may emerge when more
       pixels are used.
 \item The detector channels in a wafer have different responses due to the temperature
       gradient persistent over the wafer, which is anchored to the thermal bath only
       locally at the four sides. This will make the scaleability with the detector
       number more complex than a simple inverse law.
\end{enumerate}

We expected there would be some extra correlations between sky pixels observed at the
same by each of the pairs around $l\sim$200, but there are no features.  If we use all
of the pixels on the focal plane instead of just three, we may see some correlations.

\section{SUMMARY}\label{s5}
We presented an initial result of our end-to-end simulation of the CR noise. We
presented details of the process from the TOD to maps and power spectra using
\texttt{TOAST}. We examined the validity and scaleability of the products. We obtained
several implications to make a more robust estimate, which need to be addressed in
future studies.

\subsection*{Acknowledgments}
We acknowledge useful comments by Aritoki Suzuki, Adrian Lee, Andrea Catalano, Reijo
Keskitalo, Theodore Kisner, Giuseppe Puglisi, Hans Kristian Eriksen, and Masashi Hazumi.

Some of the results in this paper have been derived using the healpy and HEALPix
package. This research used resources of (1) JAXA super computer system JSS2 for
``Assessment of cosmic-ray for CMB observation satellite LiteBIRD'' (code name: CWU10),
(2) the National Energy Research Scientific Computing Center (NERSC), a U.S. Department
of Energy Office of Science User Facility operated under Contract No. DE-AC02-05CH11231,
(3) the Central Computing System owned and operated by the Computing Research Center at
KEK.

This work is supported in Japan by ISAS/JAXA for Pre-Phase A2 studies, by the
acceleration program of JAXA research and development directorate, by the World Premier
International Research Center Initiative (WPI) of MEXT, by the JSPS Core-to-Core Program
of A. Advanced Research Networks, and by JSPS KAKENHI Grant Numbers JP15H05891,
JP17H01115, JP17H01125, and JP18K03715. The Italian LiteBIRD phase A contribution is
supported by the Italian Space Agency (ASI Grants No. 2020-9-HH.0 and 2016-24-H.1-2018),
the National Institute for Nuclear Physics (INFN) and the National Institute for
Astrophysics (INAF). The French LiteBIRD phase A contribution is supported by the Centre
National d'E tudes Spatiale (CNES), by the Centre National de la Recherche Scientifique
(CNRS), and by the Commissariat \`{a} l'Energie Atomique (CEA). The Canadian
contribution is supported by the Canadian Space Agency. The US contribution is supported
by NASA grant no. 80NSSC18K0132.  Norwegian participation in LiteBIRD is supported by
the Research Council of Norway (Grant No. 263011). The Spanish LiteBIRD phase A
contribution is supported by the Spanish Agencia Estatal de Investigaci\'{o}n (AEI),
project refs. PID2019-110610RB-C21 and AYA2017-84185-P. Funds that support the Swedish
contributions come from the Swedish National Space Agency (SNSA/Rymdstyrelsen) and the
Swedish Research Council (Reg. no. 2019-03959). The German participation in LiteBIRD is
supported in part by the Excellence Cluster ORIGINS, which is funded by the Deutsche
Forschungsgemeinschaft (DFG, German Research Foundation) under Germany's Excellence
Strategy (Grant No. EXC-2094-390783311).

\bibliography{main} 
\bibliographystyle{spiebib} 

\end{document}